\documentclass[aps,twocolumn,epsfig
]{revtex4}
\usepackage{graphicx}
\usepackage{amsmath}
\usepackage{latexsym}
\usepackage{amsfonts}
\usepackage{amssymb}
\usepackage{array}
\usepackage{color}
\usepackage{subfigure}
\usepackage{verbatim}
\newcommand{\ket}[1]{\left\vert#1\right\rangle}
\newcommand{\bra}[1]{\left\langle#1\right\vert}

\newcommand{\RA}{\rightarrow}

\newcommand{\LRA}{\leftrightarrow}
\newcommand{\A}{\alpha}

\begin{document}
\newcommand{\Q}[1]{{\color{red}#1}}
\newcommand{\blue}[1]{{\color{blue}#1}}
\newcommand{\red}[1]{{\color{red}#1}}
\newcommand{\green}[1]{{\color{green}#1}}
\newcommand{\Change}[1]{{\color{green}#1}}
\title{
Quantum teleportation between ``particle-like'' and ``field-like'' qubits
using hybrid entanglement  under decoherence effects}

\author{Kimin Park, Seung-Woo Lee, and Hyunseok Jeong}
\affiliation{Center for Macroscopic Quantum Control,
Department of Physics and Astronomy,
Seoul National University, Seoul, 151-742, Korea}
\date{\today}

\begin{abstract}
We study quantum teleportation between two different types of optical
qubits, one of which is ``particle-like'' and the other ``field-like,''
via hybrid entangled states under the effects of decoherence.
 We find that
teleportation from particle-like to field-like qubits can be
achieved with a higher fidelity than that in the opposite direction.
However, teleportation from
field-like  to particle-like qubits is found to be more efficient in terms of the success probabilities.
Our study shows that the direction of teleportation should be considered an important factor in
developing optical hybrid architectures for quantum information processing.
\end{abstract}

\maketitle

\section{Introduction}

In optical implementations of quantum information processing (QIP), some physical
degrees of freedom of light are used for qubit encoding
\cite{PKok2007,Obrien2007,Ralph2010}.  For example, horizontal and
vertical polarization states $\ket{H}$ and $\ket{V}$ of a single
photon may be used to form a qubit basis. This type of encoding
is referred to as {\em particle-like} encoding
\cite{Ralph2010} as
individual photons are information carriers.
It is also called dual-rail encoding as it uses two distinct optical modes for a qubit \cite{KLMNature2001}.
In this type of approach, single-qubit gates
can be easily realized using linear optics elements, while two-qubit
operations are generally difficult to implement.
Alternatively, one may encode information into two
distinct states of a field mode such as the
vacuum and single photon~\cite{LundPRA2002} or two coherent states of distinct amplitudes
\cite{CochranePRA1999,
JeongPRA2002,RalphPRA2003,JeongRalphCat}.
This type of encoding is called {\em field-like} encoding (or single-rail
encoding) \cite{Ralph2010}. The coherent state encoding has
advantages for the Bell-state measurement
\cite{JeongPRA2001,Jeong2002},
and quantum computation schemes \cite{JeongPRA2002,RalphPRA2003} based on its distinctive  teleportation method
\cite{vanEnkPRA2001,JeongPRA2001}
have been developed.
Each of the two encoding schemes has its own
advantages and disadvantages for QIP \cite{LundPRL2008,ParkPRA2010}.

There have been studies on QIP based on hybrid structures using both particle-like and field-like
features of light
\cite{NemotoPRL2004,MunroNJP2005,JeongPRA2005,JeongPRA2006,LoockPRA2008,Loock2011,LeeArxiv2011}.
This type of ``hybrid architecture'' may be used to make up for the weaknesses
in both type of qubit structures.
Indeed, a near-deterministic universal quantum computation with relatively a small number of
resources is found to be possible using linear optics with a hybrid qubit composed of photon polarization and
coherent state \cite{LeeArxiv2011}. In this regard, it
is important to fully investigate such hybrid architectures, and
information transfer between different types of qubits would be a crucial task.
The quantum teleportation protocol \cite{BennettPRL1993,BouwmeesterNature1997}
can be used for such information transfer from one type of system to another.
For example, Ralph {\it et al.} discussed a scheme to perform teleportation
between a dual rail (polarization) and single rail (vacuum and single photon) qubits \cite{adaptive}.
In addition, in order to address practical conditions for such information transfer,
it would also be important to include decoherence effects caused by photon losses that are typical in optical systems.

In this paper, we study quantum teleportation between particle-like and field-like qubits under decoherence effects.
We first consider teleportation between polarization and coherent-state qubits, and that between a polarization
qubit and a qubit of the vacuum and single photon. In our study, in general, teleportation from particle-like
to field-like qubits shows higher fidelities under decoherence effects
compared to teleportation in the opposite direction.
However, teleportation from field-like to particle-like qubits is, in general, more efficient in terms of the success probabilities. This implies that the ``direction'' of teleportation should be considered to be an important factor when developing optical hybrid architectures for QIP.

This paper is organized as follows. In Sec.~II, the time
evolution of the two hybrid entangled states
under photon losses  is investigated. The degrees of
entanglement for the hybrid channels are calculated in
Sec.~III. The average fidelities and success probabilities of
teleportation are in Secs.~IV  and V.
Section~IV deals with teleportation between polarization and coherent-state qubits
while Sec.~V is devoted to investigate
 teleportation between polarization and single-rail Fock state qubits.
In Sec.~VI, the issue of postselection is discussed and investigated.
We conclude the paper with final remarks in Sec.~VII.

\section{Time evolution of teleportation channels}

The first kind of teleportation
channel considered in this paper
is a hybrid entangled state of the
photon polarization  and coherent state:
\begin{equation}
\label{eq:pc}
|\psi_{pc}\rangle=\frac{1}{\sqrt{2}}\Big(\ket{H}_p\ket{\alpha}_c+\ket{V}_p\ket{-\alpha}_c\Big),
\end{equation}
where $|\pm\alpha\rangle$ are coherent states of amplitudes $\pm\alpha$.
We assume that $\alpha$ is real for simplicity throughout the paper without loss of generality.
The other one is a hybrid channel of  the photon polarization and the single-rail photonic qubit
\begin{equation}
\label{eq:ps}
\ket{\psi_{ps}}=\frac{1}{\sqrt{2}}\Big(\ket{H}_p\ket{0}_s+\ket{V}_p\ket{1}_s\Big),
\end{equation}
where $\ket{0}$ and $\ket{1}$ denote the vacuum and the single
photon state in the Fock basis, respectively, comprising a
field-like (single-rail) qubit. Here, $p$, $c$ and $s$ respectively stand for
polarization, coherent state and single-rail Fock
state qubits.
It is known that the hybrid channel $|\psi_{pc}\rangle$
can in principle be produced using a weak cross-Kerr nonlinear
interaction between a polarization (dual-rail) single photon qubit
and a coherent state \cite{GerryPRA1999,NemotoPRL2004,JeongPRA2005}.
However, it is highly challenging to perform the required nonlinear
interaction with high efficiency \cite{Shapiro2006,Shapiro2007,Banacloche2010}. The hybrid
channel $|\psi_{ps}\rangle$ can be generated using a parametric down conversion, a Bell
state measurement with polarization qubits and an adaptive
measurement \cite{adaptive}.

We consider decoherence caused by
photon loss (dissipation) on the teleportation channels. The dissipation
for state $\rho$ is
described by the master equation under the Born-Markov approximation
with zero temperature environment~\cite{PhoenixPRA1990}
\begin{equation}
\label{eq:me}
 \frac{\partial \rho}{\partial\tau}=\hat{J}\rho+\hat{L}\rho,
\end{equation}
where $\tau$ is the system-bath interaction time.
Lindblad superoperators $\hat{J}$ and $\hat{L}$ are defined as
$\hat{J}\rho=\gamma\sum_ia_i \rho a_i^\dagger$ and
$\hat{L}\rho=-\sum_i \gamma(a_i^\dagger a_i \rho + \rho
a_i^\dagger a_i)/2$, where $\gamma$ is the decay constant determined by
the coupling strength of the system and environment,
and $a_i$ is the annihilation operator for  mode $i$.
Throughout this paper, we assume that the decay constant $\gamma$ is same for modes $p$, $c$ and $s$, {\it i.e.},
the photon loss occurs at the same rate for all modes.

The formal solution of Eq.~(\ref{eq:me}) is written as
$\rho(\tau)={\rm exp}  [(\hat{J}+\hat{L})\tau]\rho(0)$, where
$\rho(0)$ is the initial density operator at $\tau=0$. By solving this equation
we obtain the decohered density matrix for the initial state of the hybrid channel
$|\psi_{pc}\rangle$
in Eq.~(\ref{eq:pc}) as
\begin{align}
\label{eq:dissipation}
\rho_{pc} (t;\A)&=\frac{1}{2}\Big[\big\{t^2\ket{H}_p\bra{H}+(1-t^2)\ket{0}_p\bra{0}\big\}\otimes\ket{t\alpha}_c\bra{t\alpha}
\nonumber\\
&+\big\{t^2\ket{V}_p\bra{V}+(1-t^2)\ket{0}_p\bra{0}\big\}\otimes\ket{-t\alpha}_c\bra{-t\alpha}
\nonumber\\
&+t^2 Q(t)\big(\ket{H}_p\bra{V}\otimes\ket{t\alpha}_c\bra{-t\alpha}+h.c.\big)\Big]
\end{align}
where the parameter $t = e^{-\gamma\tau /2}$ describes the amplitude
decay, and $Q(t)\equiv e^{-2\alpha^2 (1-t^2)}$
reflects the reduction of
the  off-diagonal coherent-state dyadic $\ket{\A}\bra{-\A}$ and its hermitian conjugate.
We define the normalized time as $r= (1- t^2 )^{1/2}$ which gives a value $r=0$ at $\tau=0$ and $r=1$ at $\tau=\infty$. Likewise, we obtain the decohered density
matrix $\rho_{ps}(t)$ for the initial state in the channel in Eq.~(\ref{eq:ps}) as
\begin{align}
\label{eq:paevolution}
&\rho_{ps}(t)=\frac{1}{2}\Big[\big\{t^2\ket{H}_p\bra{H}+(1-t^2)\ket{0}_p\bra{0}\big\}\otimes\ket{0}_s\bra{0}\nonumber\\
&+\big\{t^2\ket{V}_p\bra{V}+(1-t^2)\ket{0}_p\bra{0}\big\}\otimes\big\{t^2\ket{1}_s\bra{1}\nonumber\\
&+(1-t^2)\ket{0}_s\bra{0}\big\}+t^3\big(\ket{H}_p\bra{V}\otimes\ket{0}_s\bra{1}+h.c.\big)\Big].
\end{align}
As shown in Eqs.~(\ref{eq:dissipation}) and
(\ref{eq:paevolution}), photon loss induces (i)
the decay of the amplitude of coherent state as $\ket{\A}\RA\ket{t\A}$, (ii) the
transition of the polarization states $\ket{H}_p\bra{H}$ and $\ket{V}_p\bra{V}$ into vacuum state
$\ket{0}_p\bra{0}$, which causes an escape error out of the qubit
space, (iii) the transition of the single photon Fock state $\ket{1}_s\bra{1}$ into
vacuum state $\ket{0}_s\bra{0}$, a flip error of the qubit,
and (iv) the decrease of the coefficients of coherence (off-diagonal) terms with $t^2
Q(t)$ in Eq.~(\ref{eq:dissipation}) and $t^3$ in
Eq.~(\ref{eq:paevolution}).

\section{Entanglement of hybrid channels}

\begin{figure}[tbp]
\includegraphics[width=250px]{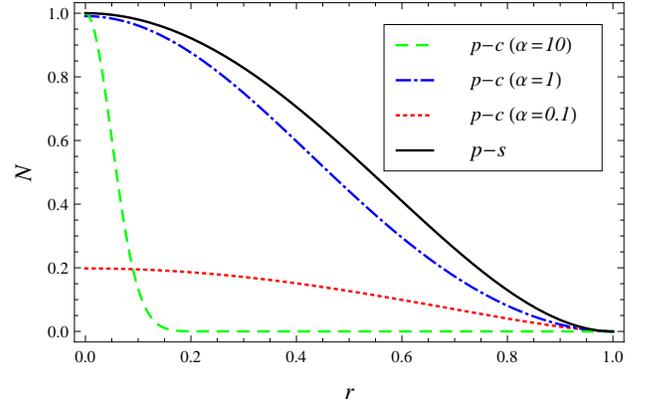}
\caption{ (Color online) Negativity $N$ of
the hybrid channels, $\rho_{pc}$ (dotted, dot-dashed, and
dashed curves) and $\rho_{ps}$ (solid curve) in Eqs.~(\ref{eq:dissipation}) and (\ref{eq:paevolution}), against the normalized time $r$
under decoherence.} \label{fig:negativity}
\end{figure}

The negativity of state $\rho$, known as a measure of entanglement,
is defined as \cite{ZyczkowskiPRA1998,VidalPRA2002}
\begin{equation}
\label{eq:negativity}
N(\rho)=||\rho^{T_B}||-1=2\sum_{\lambda_i<0}{|\lambda_i|},
\end{equation}
where $\rho^{T_B}$ is the partial transpose of $\rho$ about one mode
of composite system (say mode $B$ here), $||\cdot||$ denotes the
trace norm and $\lambda_i$'s are negative eigenvalues of
$\rho^{T_B}$.
We calculate the negativity of the decohered channel $\rho_{pc}$
given in Eq.~(\ref{eq:dissipation}) as
\begin{align}
\label{eq:negativitypC}
&N(\rho_{pc}(t;\A))=\frac{t^2}{2N_+^2 N_-^2} \Big\{(Q(t)-1 )(N_+^2+N_-^2)\nonumber\\
&~~~~~+\sqrt{16 Q(t) N_+^2 N_-^2+(1 - Q(t)  )^2(N_+^2+N_-^2)^2}\Big\}
\end{align}
where
$N_\pm=(2\pm2e^{-2t^2\A^2})^{-1/2}$ are normalization factors for equal superpositions of coherent states
$\ket{\pm}=N_\pm(\ket{t\A}\pm\ket{-t\A})$. This is obtained by representing the coherent state qubit part of Eq.~(\ref{eq:dissipation}) in the orthogonal basis $\{\ket{\pm}\}$ and performing calculations following
Eq.~(\ref{eq:negativity}).
The negativity of the decohered entangled channel $\rho_{ps}$ in
Eq.~(\ref{eq:paevolution}) is also obtained as
\begin{align}
\label{eq:negativityps}
N(\rho_{ps})=t^4.
\end{align}
The degrees of entanglement for the two channels are plotted in Fig.~\ref{fig:negativity}, and we find
that entanglement contained in $\ket{\psi_{ps}}$ is more
robust to decoherence than that of $\ket{\psi_{pc}}$.
Obviously,  state
$\ket{\psi_{pc}}$ is more entangled  when $\alpha$ is larger at the initial time.
As $\A\rightarrow0$, the initial state
approaches a product state with no entanglement.
However, when the initial value of $\alpha$ is larger, the slope of the decrease of entanglement is steeper,
{\it i.e.}, entanglement  disappears more  rapidly. The reason for this is  that state $\ket{\psi_{pc}}$  becomes a more ``macroscopic'' quantum superposition, fragile to decoherence, when $\alpha$ is large.
This  feature has been pointed out in a number of previous studies \cite{KimPRA1992,JeongPRA2001,Wilson2002,JLKPRA2000,JParkPRA2012} with various versions of
continuous-variable superpositions and entangled states.
In our case, when $\A\approx
1$, entanglement seems most robust to decoherence considering both the initial value and
the decrease slope of entanglement.

\section{Teleportation between polarization and coherent-state qubits}

We now consider quantum teleportation using the hybrid channels.
Besides the hybrid channels, a Bell-state measurements and single-qubit unitary
transforms, $\sigma_x$ and $\sigma_z$ operations, at the receiver's site are required to complete
the teleportation process.
In order to avoid unrealistic assumptions,
we assume throughout the paper that only linear optics elements are available
besides the hybrid quantum channels.

In this Section, we first investigate quantum teleportation between
polarization and coherent state qubits through the decohered entangled
state $\rho_{pc}$ in Eq.~(\ref{eq:dissipation}). For 
convenience, we use the arrow $A\RA B$ for the teleportation from
qubit type $A$ to  type $B$ when a hybrid
entangled state composed of two qubits with types $A$ and $B$ is
used as the teleportation channel. For example, $p\RA c$ indicates
teleportation from polarization to coherent-state qubits, and
$c\RA p$ vice versa.

\subsection{Teleportation fidelities}

The teleportation fidelity $F$ is defined as
$F=\bra{\psi_\mathrm{t}}\rho_\mathrm{out}\ket{\psi_\mathrm{t}}$
where $\ket{\psi_\mathrm{t}}$ is the target state of teleportation
and $\rho_\mathrm{out}$ is the density operator of the output qubit.
Due to the non-orthogonality of two coherent states, it is not trivial to define
the fidelity between a polarization qubit and a coherent-state qubit.
In the case of teleportation
from a polarization qubit, $\ket{\psi_\mathrm{t}}_p=a\ket{H}_p+b\ket{V}_p$, to a coherent state qubit,
it would be reasonable to choose the target state as
\begin{eqnarray}
\label{eq:targetc}
\ket{\psi_\mathrm{t}}_c=N(a\ket{t\A}_c+b\ket{-t\A}_c),
\end{eqnarray}
where $N=\{1+(a b^*+a^* b)e^{-2t^2\A^2}\}^{-1/2}$ is the normalization
factor.
We note that we take a dynamic qubit basis $\{|\pm t\alpha\rangle \}$ in order to reflect the decrease of the amplitude under photon losses
\cite{JeongPRA2001}, and that $t$ is considered a known value. Conversely, for the teleportation of
opposite direction ($c \RA p$) the state in Eq.~(\ref{eq:targetc})
is considered the input qubit and
$\ket{\psi_\mathrm{t}}_p=a\ket{H}_p+b\ket{V}_p$
the target state.

The Bell-state measurement, an essential part of quantum teleportation, discriminates
four Bell states:
\begin{align}
&\ket{B_{1,2}}_{pp'}=\frac{1}{\sqrt{2}}(\ket{H}_{p}\ket{H}_{p'}\pm\ket{V}_{p}\ket{V}_{p'}),\\
&\ket{B_{3,4}}_{pp'}=\frac{1}{\sqrt{2}}(\ket{H}_{p}\ket{V}_{p'}\pm\ket{V}_{p}\ket{H}_{p'}).
\end{align}
The Bell-state
measurement in polarization modes can be performed by a 50:50
beam splitter, two polarizing beam splitters and photon detectors
\cite{LutkenhausPRA1999}, which discriminates only $\ket{B_{3,4}}_{pp'}$ successfully.
The net effect of this process is equivalent to taking the inner
product of the total density matrix $\ket{\psi_\mathrm{t}}_p
\bra{\psi_\mathrm{t}}\otimes \rho_{p'c}(t;\alpha)$ with a Bell state,
and an appropriate unitary transform is applied to reconstruct the original state.
For example, when one of the Bell states,
$\ket{B_1}_{p p'}$, is
measured, the output state for the teleportation from a polarization
to a coherent state qubit for an input state
$\ket{\psi_\mathrm{t}}_p$ is given as
\begin{equation}
\label{eq:pcmiddle}
\rho^{p\RA c}_\mathrm{out}=\frac{_{pp'}\bra{B_1}\big\{\ket{\psi_\mathrm{t}}_p \bra{\psi_\mathrm{t}}\otimes \rho_{p'c}(t;\alpha)\big\}\ket{B_1}_{pp'}}{\mathrm{Tr}\Big[\ket{B_1}_{pp'}\bra{B_1}\big\{\ket{\psi_\mathrm{t}}_p \bra{\psi_\mathrm{t}}\otimes \rho_{p'c}(t;\alpha)\big\}\Big]}.
\end{equation}
In this case, no unitary transform is required.
In the cases of the other outcomes, the required unitary transforms for the coherent state part are
\begin{align}
\label{eq:unitary}
&Z_{c}:\ket{\pm t\alpha}_{c}\RA\pm\ket{\pm t\alpha}_{c},\nonumber\\
&X_{c}:\ket{\pm t\alpha}_{c}\RA\ket{\mp t\alpha}_{c},
\end{align}
after which the state of Eq.~(\ref{eq:pcmiddle}) is obtained.
One or both of these operations should be applied depending on the Bell-state measurement  outcome \cite{JeongPRA2001}.
It is relatively easy to perform $X_{c}$ using phase shifter, while the implementation of
$Z_{c}$ is non-trivial \cite{JeongPRA2001,LundPRL2008}.
The displacement operation can approximate the $Z_{c}$ operation
\cite{JeongPRA2001,JeongPRA2002} but it becomes effective only for $\alpha\gg1$.

Therefore, we  slightly modify the Bell measurement to obtain different success outcomes (instead of $\ket{B_{3,4}}_{pp'}$) to avoid $Z_{c}$ operation. A  Hadamard operation on the first mode, and a bit flip operator $X$ with a Hadamard operation on the second input mode in $\{\ket{H},\ket{V}\}$ basis respectively transforms the Bell states to
\begin{align}
&\ket{B_{1}}_{pp'}\rightarrow\ket{B_{3}}_{pp'},&\ket{B_{2}}_{pp'}\rightarrow\ket{B_{1}}_{pp'},\\
&\ket{B_{3}}_{pp'}\rightarrow\ket{B_{4}}_{pp'},&\ket{B_{4}}_{pp'}\rightarrow\ket{B_{2}}_{pp'}.
\end{align}
 When the Bell measurement setup is applied after this transformation, we can discriminate the initial states $\ket{B_1}_{pp'}$ and $\ket{B_3}_{pp'}$ before the above transformation as successful outcomes. In this way,  only the $X_{c}$ operation is necessary on the output state of teleportation.

Inserting explicit forms of $\rho_{p'c}(t;\alpha)$ in Eq.~(\ref{eq:dissipation}) and $\ket{\psi_\mathrm{t}}_p=a\ket{H}_p+b\ket{V}_p$ into (\ref{eq:pcmiddle}) gives
\begin{widetext}
\begin{align}
\label{eq:pCout} \rho^{p\RA c}_\mathrm{out}= \frac{|a|^2
\ket{t\A}_c\bra{t\A}+|b|^2 \ket{-t\A}_c\bra{-t\A}+Q(t) \big(a b^*
\ket{t\A}_c\bra{-t\A}+
a^* b\ket{-t\A}_c\bra{t\A}
\big)}{1+e^{-2\A^2}(a b^*+a^* b)}.
\end{align}
We find the fidelity between the output state $\rho^{p\RA c}_\mathrm{out}$ in
Eq.~(\ref{eq:pCout}) and the target state
$\ket{\psi_\mathrm{t}}_c=N(a \ket{t\A}_c+b\ket{-t\A})_c$ as
\begin{align}
\label{eq:Fspecific}
F_{p\RA c}=~_c\bra{\psi_\mathrm{t}}\rho^{p\RA c}_\mathrm{out}\ket{\psi_\mathrm{t}}_c=
\frac{|a|^2
|a+b S|^2+|b|^2 |a S+b|^2+2Q(t) \mathrm{Re}\big[a b^*
(a+b S)(a^* S+b^*)\big]}{N^{-2}\Big\{1+e^{-2\A^2}(a b^*+a^* b)\Big\}}
\end{align}
\end{widetext}
where $S=\langle t\A|-t\A\rangle=\exp(-2t^2\A^2)$ is the overlap between the dynamic qubit basis states.

We now find the average teleportation fidelity over all possible input states.
For convenience, we parametrize the unknown values of the input state
as
$a=\cos[\theta/2]\exp[\mathrm{i}
\phi/2]$ and $b=\sin[\theta/2]\exp[-\mathrm{i}
\phi/2]$, where $0\leq \phi<2\pi$ and $0\leq\theta<\pi$.
The average of $F_{p\rightarrow c}(\theta,\phi)$ in Eq.~(\ref{eq:Fspecific}) over all input states
is then obtained
using Eq.~(\ref{eq:Gdef}) as
\begin{align}
\label{eq:explicitav}
&F_{p\RA c}(t)=\langle F_{p\rightarrow c}(\theta,\phi)\rangle_{\theta,\phi}
\nonumber\\&~~~~=\frac{1}{4\pi}\int_0^\pi d\theta\sin\theta\int_0^{2\pi}d\phi  F_{p\rightarrow c}(\theta,\phi)\nonumber\\
&~~~~=\frac{Q(t)}{Q(t)-1}\Big\{2G[|a|^4]+\big(2S^2+2Q(t))G[|a|^2|b|^2]\nonumber\\
&+(S Q(t)+S) G[a b^*+a^*b]+S^2 Q(t) G[a^2 {b^*}^2+{a^*}^2 b^2]\Big\},
\end{align}
where $G[f]$ for arbitrary value or function $f=f(\theta,\phi)$ is
\begin{align}
\label{eq:Gdef}
G[f]
=\Big\langle \frac{f}{1+Q(t)S(a b^*+a^*b)}-\frac{f}{1+S(a b^*+a^*b)}\Big\rangle_{\theta,\phi}
\end{align}
with
\begin{align}
&\Big\langle \frac{|a|^4}{1+x(a b^*+a^*b)}\Big\rangle_{\theta,\phi}=~\frac{x+\frac{-1+3x^3}{\tanh[x]}}{8x^3},\\
&\Big\langle \frac{|a|^2|b|^2}{1+x(a b^*+a^*b)}\Big\rangle_{\theta,\phi}=-\frac{x+\frac{-1-x^2}{\tanh[x]}}{8x^3},\\
&\Big\langle \frac{a b^*+a^*b}{1+x(a b^*+a^*b)}\Big\rangle_{\theta,\phi}=~\frac{1}{x}-\log[\frac{1+x}{1-x}]\frac{1}{2x^2},\\
&\Big\langle \frac{a^2 {b^*}^2+{a^*}^2 b^2}{1+x(a b^*+a^*b)}\Big\rangle_{\theta,\phi}=\frac{2-x^2}{4x^3}\log[\frac{1+x}{1-x}]-\frac{1}{x^2}
\end{align}
for arbitrary value $x$ independent of $\theta$ and $\phi$.

Now, we consider teleportation from a
coherent state qubit to a polarization qubit. The Bell-state
measurement for coherent-state qubits can be performed using a 50:50
beam splitter and two photon number parity measurements
\cite{JeongPRA2001}.
The input qubit of the form of Eq.~(\ref{eq:targetc}) together with the coherent-state part
of channel $\rho_{pc'}(t;\A)$ passes through the 50:50 beam splitter and evolves as
\begin{align}
\label{eq:beam1}
&\big(a\ket{\beta}+b\ket{-\beta}\big)_c\ket{\beta}_{c'}\RA a|\sqrt{2}\beta\rangle_c\ket{0}_{c'}+b\ket{0}_{c}|\sqrt{2}\beta\rangle_{c'}\nonumber\\
&\big(a\ket{\beta}+b\ket{-\beta}\big)_{c}\ket{-\beta}_{c'}\RA a\ket{0}_{c}|-\sqrt{2}\beta\rangle_{c'}+b|-\sqrt{2}\beta\rangle_{c}\ket{0}_{c'},
\end{align}
where $\beta=t\A$.
We note that the photons move to either of the two modes so that only one of the two detectors
can detect any photon(s).
The projection operators $O_j$ for the outcomes $j$ of the two parity measurements
can then be written as
\begin{eqnarray}
\label{eq:O1}
\hat O_1&=&\sum_{n=1}^\infty |2n\rangle_A\langle2n|\otimes|0\rangle_B\langle0|,\\
\hat O_2&=&\sum_{n=1}^\infty |2n-1\rangle_A\langle2n-1|\otimes|0\rangle_B\langle0|,\\
\hat O_3&=&\sum_{n=1}^\infty |0\rangle_A\langle0|\otimes|2n\rangle_B\langle2n|,\\
\hat O_4&=&\sum_{n=1}^\infty |0\rangle_A\langle0|\otimes|2n-1\rangle_B\langle2n-1|,
\label{eq:O4}
\end{eqnarray}
where subscripts 1, 2, 3 and 4 represent four Bell states
\begin{eqnarray}
&|{\cal B}_{1,2}\rangle_{cc'}\propto|\alpha\rangle_c|\alpha\rangle _{c'}\pm|-\alpha\rangle_{c}|-\alpha\rangle_{c'}, \\
&|{\cal B}_{3,4}\rangle_{cc'}\propto|\alpha\rangle_c |-\alpha\rangle_{c'}\pm|-\alpha\rangle_{c}|\alpha\rangle_{c'},
\end{eqnarray}
respectively.
In addition, the error projection operator
$\hat O_\mathrm{e}=|0\rangle_A\langle0|\otimes|0\rangle_B\langle0|$
should also be considered because there is possibility for both the detectors not to register anything,
even though such probability approaches zero for $\alpha\gg1$.

In the calculation to obtain the output density matrix when the element of the parity measurement $\hat{O}_1$ is measured, the terms such as $\ket{0}_{c}\ket{\sqrt{2}\beta}_{c'}$ and $\ket{0}_{c}\ket{-\sqrt{2}\beta}_{c'}$ in Eq.~(\ref{eq:beam1}) are
erased from the resultant density matrix due to the orthogonality of vacuum state in these terms and non-zero number states contained in $\hat{O}_1$. Other terms form the same factor $\sum_{n=1}^\infty \langle 2n|\sqrt{2}\beta\rangle\langle \pm\sqrt{2}\beta|2n\rangle=\cosh(2\beta^2)-1$, which is factored out into the normalization factor. When $\hat{O}_2$, $\hat{O}_3$ and $\hat{O}_4$  are measured in the parity measurements, the unitary transforms required are Pauli matrices $(\sigma_z)_p$, $(\sigma_x)_p$ and $(\sigma_y)_p$ in the basis set of $\{\ket{H},\ket{V}\}$, respectively. 

  The overall effect of the Bell-state measurement and unitary transform is found to be replacement of $\ket{t\A}_{c'}$ ($\bra{t\A}_{c'}$) with $a$ ($a^*$) and $\ket{-t\A}_{c'}$ ($\bra{-t\A}_{c'}$) with $b$ ($b^*$) in the teleportation channel $\rho_{pc'}(t;\A)$ in Eq.~(\ref{eq:dissipation}). We obtain after the normalization
\begin{align}
\label{eq:Cpout}
\rho^{c\RA p}_\mathrm{out}=&\frac{\mathrm{Tr}_{cc'}\big[(\hat{O}_1 U_\mathrm{BS})_{cc'}(\rho_{pc'}(t;\alpha)\otimes\ket{\psi_\mathrm{t}}_c \bra{\psi_\mathrm{t}})(U^\dagger_\mathrm{BS})_{cc'}\big]}{\mathrm{Tr}\Big[(\hat{O}_1 U_\mathrm{BS})_{cc'}(\rho_{pc'}(t;\alpha)\otimes\ket{\psi_\mathrm{t}}_c \bra{\psi_\mathrm{t}})(U^\dagger_\mathrm{BS})_{cc'}\Big]}\nonumber\\
=&t^2|a|^2 \ket{H}_p\bra{H}+t^2|b|^2 \ket{V}_p\bra{V}+(1-t^2)\ket{0}_p\bra{0}\nonumber\\
&+t^2 Q(t) \big(a b^* \ket{H}_p\bra{V}+a^* b \ket{V}_p\bra{H}\big),
\end{align}
where $U_\mathrm{BS}$ represents the beam splitter operator.
The fidelity is then
\begin{align}
\label{eq:CpF}
F_{c\RA p}(\theta,\phi)&={}_p\bra{\psi_\mathrm{t}}\rho^{c\RA p}_\mathrm{out}\ket{\psi_\mathrm{t}}_p
\nonumber\\
&=t^2\big(|a|^4
+|b|^4+Q(t) |a|^2 |b|^2\big)
\end{align}
and its average can be calculated in a similar way as Eq.~(\ref{eq:explicitav})
\begin{equation}
\label{eq:Cpav} F_{c\RA p}(t)=t^2
\bigg(\frac{2}{3}+\frac{Q(t)}{3}\bigg).
\end{equation}

\begin{figure}[tbp]
\subfigure[~$\A=0.1$]{
\includegraphics[width=115px]{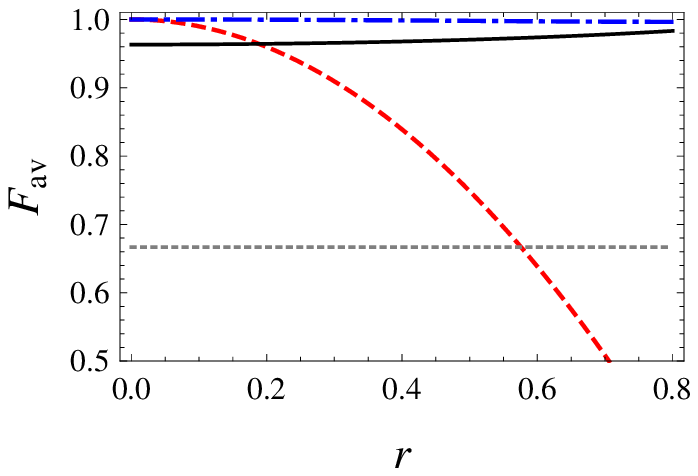}
}
\subfigure[~$\A=1$ ]{
\includegraphics[width=115px]{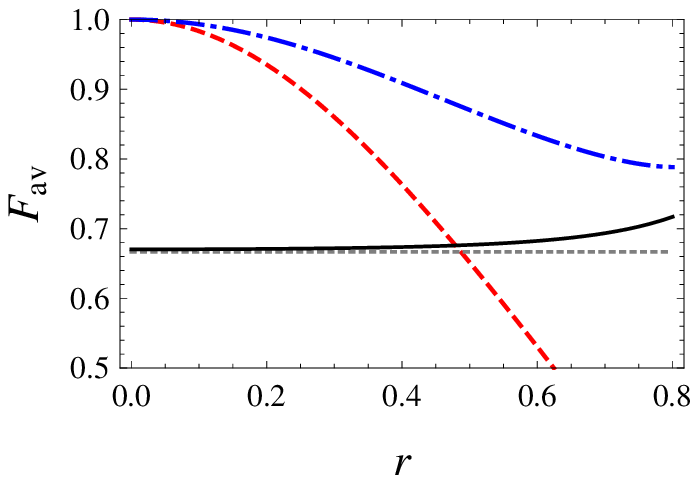}
}
\subfigure[~$\A=2$ ]{
\includegraphics[width=115px]{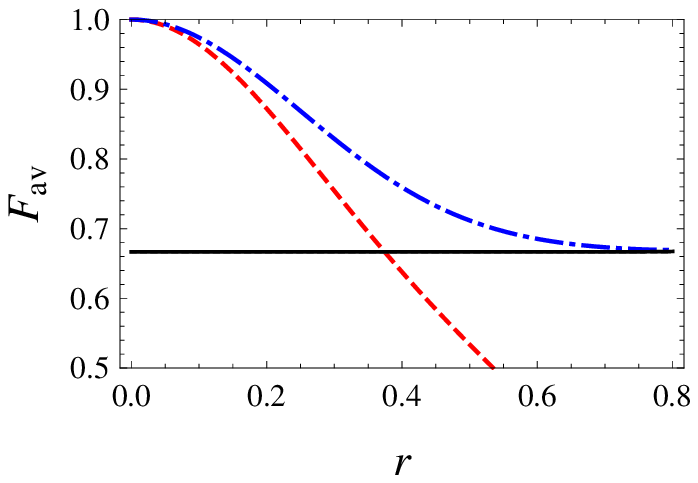}
}
\subfigure[~$\A=10$]{
\includegraphics[width=115px]{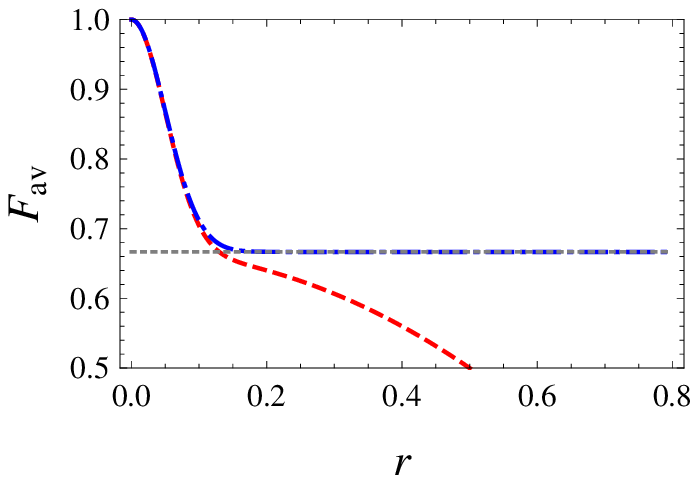}
} \caption{ (Color online) Average fidelities of teleportation
from polarization to coherent state qubits ($p\RA c$, dot-dashed curves)
and of teleportation in the opposite direction ($c\RA
p$, dashed curves) for several values of $\A$. The classical limits $F^{p\RA c}_{cl}$
 (solid lines) for $p\RA c$  and $2/3$  (dotted lines)
for  $c\RA p$ are plotted for comparison. } \label{fig:2DpC}
\end{figure}

We also consider the classical limits of teleportation fidelity for comparison.
Here, a classical limit means the maximum average fidelity of teleportation (disembodied transport of an unknown quantum state) by means of a classical communication channel without any entanglement.
It is well known that the classical limit of fidelity for teleporting a qubit
using the standard teleportation protocol is 2/3 \cite{MassarPRL1995}.
It can be directly applied when teleporting a coherent-state qubit to a polarization qubit ($c\RA p$).
However, due to the nonorthogonality of two coherent states,
the classical limit for teleportation from polarization  to coherent-state qubits
($p\RA c$) is larger than $2/3$.
A simple way to consider the classical limit is as follows.
The optimal output state of teleportation without quantum entanglement is
$\rho_{cl}=|a|^2\ket{t\A}\bra{t\A}+|b|^2\ket{-t\A}\bra{-t\A}$,
where the amplitude decay was also considered. This state is obtained by preparing either state $\ket{t\A}\bra{t\A}$ or state $\ket{-t\A}\bra{-t\A}$ depending on the measurement outcomes
of the input state.
The average fidelity with the target state $\ket{\psi_\mathrm{t}}_c$ is
\begin{align}
F^{p\RA c}_{cl}(t)&=\Big\langle\bra{\psi_t}\rho_{cl}\ket{\psi_t}\Big\rangle_{\theta,\phi}\nonumber\\
&=\frac{S+3S^3-(S^4-1)}{4S^3}\mathrm{sinh}^{-1}\Big[\frac{S}{\sqrt{1-S^2}}\Big].
\end{align}
In the limit of $\alpha\rightarrow \infty$ where the two basis coherent  states become
orthogonal, $F^{p\RA c}_{cl}(t)$ approaches 2/3.

In Fig.~\ref{fig:2DpC}, we plot the time evolution of average
teleportation fidelities for different coherent state amplitudes
$\alpha=0.1,1,2,10$ against the normalized time $r$.
When $\A$ is large, the
teleportation fidelities of both directions $p\LRA c$ decrease
rapidly down to the classical limit
after short time as shown in Fig.~\ref{fig:2DpC}(d).
This result is in agreement with the fast decay of entanglement in the
channel presented in Fig.~\ref{fig:negativity}.
When $\A$ is relatively small, the
average fidelity for the teleportation $p\LRA c$ is close to 1 in spite of the small amount
of entanglement in the channel shown in Fig.~\ref{fig:negativity}.
This can be attributed to the effect of non-orthogonality
between coherent states $\ket{t\A}$ and $\ket{-t\A}$.

In our analysis, as implied in Fig.~\ref{fig:2DpC},
the fidelity of teleportation from polarization to
coherent state qubit~($p\RA c$) is shown to be always larger than that of
teleportation in the opposite direction~($c\RA p$). In the region over
the classical limit $2/3$, the gap between these two fidelities for
a given $r$ decreases as $\A$ becomes larger as shown in
Fig.~\ref{fig:2DpC}.
This gap can be obtained and explained as follows.
In the limit of large $\A$,
the output state of the teleportation ($p\RA c$) in
Eq.~(\ref{eq:pCout}) can be approximated as
\begin{align}
\label{eq:largeA}
\rho^{p\RA c}_\mathrm{out}\approx &|a|^2 \ket{t\A}_c\bra{t\A}+|b|^2
\ket{-t\A}_c\bra{-t\A}\nonumber\\
&+Q(t) (a b^*\ket{t\A}_c\bra{-t\A}+a^* b\ket{-t\A}_c\bra{t\A}).
\end{align}
The comparison between
the output state for $p\RA c$ in Eq.~(\ref{eq:largeA}) and the output state for $c\RA p$ in Eq.~(\ref{eq:Cpout})  implies that the
difference between $F_{c\rightarrow p}$ and $F_{p\rightarrow c}$ for large values of $\alpha$ can be attributed to the term $(1-t^2)\ket{0}_p\bra{0}$
in Eq.~(\ref{eq:Cpout}). The
fidelity between the output state in Eq.~(\ref{eq:largeA}) and the
target state $\ket{\psi_\mathrm{t}}_c$ is $|a|^4 +|b|^4 + 2
Q(t) |a| |b|$ and its average is calculated to be $(2+Q(t))/3$. By
subtracting Eq.~(\ref{eq:Cpav}) from this, we obtain the
gap between the two fidelities as $(1-t^2)(2+Q(t))/3$.
In the limit of $\alpha\rightarrow\infty$, the gap at time $t_{cl}$ that satisfies
$F_{c\RA p}(t_{cl})=2/3$
approaches zero.

In further detail, the difference between $F_{p\RA c}$ and
$F_{c\RA p}$
 can be explained by two effects: (i) the overlap
between $\ket{t\A}$ and $\ket{-t\A}$ which is
dominant at the region $t\A\ll 1$, and (ii)
the effect that the polarization qubit turns into the
vacuum state by photon loss so that the output can no longer be in the original qubit space:
this is not the case for the dynamic qubit basis using
 $\ket{\pm t\A}$.
In the case of $p\RA c$, the vacuum introduced by photon loss  is detected
during the Bell-state measurement and discarded by virtue of its particle-like
nature. This filtering effect in the Bell-state measurement for the polarization qubits
enhances the fidelity  $F_{p\RA c}$
over $F_{c\RA p}$.
While the average fidelity $F_{p\RA c}$
is always higher than the classical limit,
$F_{c\RA p}$ degrades below
its classical limit because of the vacuum component in the output state.

\subsection{Success probabilities}

\begin{figure}[tbp]
\includegraphics[width=250px]{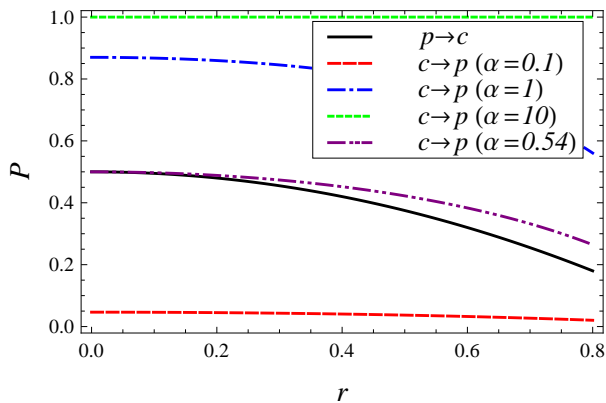}
\caption{ (Color online) Success probability for teleportation
between polarization and coherent qubits for different coherent
state amplitudes ($\A = 0.1, 1, 0.54, 10$) against the normalized
time $r$ under decoherence.} \label{fig:PpC}
\end{figure}

An event of the teleportation process should be discarded either when the Bell-state measurement fails or
when the appropriate unitary transform is unavailable.
Due to these discarded events, the success probability of the teleportation process becomes smaller than  unity.
We first consider the teleportation of $p\RA c$.
The Bell-state measurement for the teleportation of $p\RA c$ is to distinguish the four Bell states
of polarization qubits.
This type of Bell-state measurement can identify only two of the four Bell states
\cite{LutkenhausPRA1999}. As briefly explained in the previous subsection, the choice of
two successful outcomes can be made arbitrary with a few simple gate operations.
%
We here take $\ket{B_1}_{p p'}$ and $\ket{B_3}_{p p'}$ as successful outcomes and discard  the other results.
Considering these inherent limitations,
the success probability  of  teleportation $p\RA c$ cannot exceed 1/2.
Beside these, a failure of the Bell-state measurement also occurs when
the photon is lost from the channel in the polarization qubit part.
Such loss can be immediately noticed at the detectors used for the Bell-state measurement and should be considered
for the success probability.

The success probability for a specific input state is $P(\theta,\phi)=\mathrm{Tr}\big[(\ket{B_1}_{p p'}\bra{B_1}+\ket{B_3}_{p p'}\bra{B_3})\{\ket{\psi_\mathrm{t}}_p
\bra{\psi_\mathrm{t}}\otimes \rho_{p'c}(t;\alpha)\}\big]$. In fact, the explicit form of $P(\theta,\phi)$ is
obtained during the normalization of the output state $\rho_{\rm out}^{p\rightarrow c}$ as the inverse of the normalization factor
as implied in  Eqs.~(\ref{eq:pcmiddle}) and (\ref{eq:pCout}).
We find $P_{p\RA c}(\theta,\phi)=t^2(1+A\sin\theta \cos\phi)/2$ and the total success probability over all of the input states can be calculated by
\begin{align}
 P_{p\RA c}(t)=&\langle P_{p\RA c}(\theta,\phi)\rangle_{\theta,\phi}
=\frac{t^2}{2}, \label{eq:Ppc}
\end{align}

On the other hand, teleportation for $c\RA p$ can be performed with a high probability close to unity only using linear optics. This is due to the two reasons as follows.
First, the Bell-state measurement for the coherent-state qubits, required for the sender's site in this process, can discriminate between
all four Bell states \cite{JeongPRA2001}.
Second, the single-qubit unitary transforms for the polarization qubit, to be performed in the receiver's site,
are straightforward for any outputs.
The results are discarded only when
no photons are detected in the Bell-state measurement.
Of course, when loss caused by decoherence occurs,
the parity measurement scheme used for the
Bell-state measurements in the coherent-state basis
cannot filter out ``wrong results'' in the polarization part,
which is obviously different from the Bell-state measurement with polarization qubits,
and this type of errors will be reflected
in the degradation of the fidelity.

  The success probability of $c\RA p$ teleportation for a given input state
is then obtained by
\begin{equation}
\label{eq:Pcp}
P_{c\RA p}(\theta,\phi)=\sum_{i}\langle U_\textrm{BS}^\dagger\hat{O}_i U_\textrm{BS}\rangle
=(1-S)/(1+S\sin\theta \cos\phi)
\end{equation}
where $\hat{O}_i$'s
  are the
projection operators introduced in the previous subsection and $U_\textrm{BS}$ is the operator for 50:50 beam splitter.  The success
probability of all input states can be calculated in the same way described above and the result is
\begin{equation}
\label{eq:PCp}
 P_{c\RA p}(t)=\frac{S^{-1}-1}{2}\log\bigg(\frac{1+S}{1-S}\bigg).
\end{equation}

The success probabilities in Eqs.~(\ref{eq:Ppc}) and (\ref{eq:PCp}) are plotted and compared for
several values of $\A$ in
Fig.~\ref{fig:PpC}. The success probability $P_{p\RA c}(t)$ is
invariant under the change of $\A$, while $P_{c\RA p}(t)$ becomes
larger as $\A$ increases. As the hybrid channel undergoes decoherence,
both  $P_{p\RA c}(t)$  and  $P_{c\RA p}(t)$  decrease due to photon
losses. The decrease of $P_{c\RA p}(t)$ becomes negligible
for $\A\gg1$ as the proportion of the vacuum state in the coherent state is very small.
When $\A\approx 0.54$, probabilities $P_{p\RA c}(t)$  and  $P_{c\RA p}(t)$ become comparable
for all ranges of $r$.

\section{Teleportation between polarization and single-rail Fock state qubits}

In this Section, we go on to investigate teleportation between polarization
and single-rail Fock state qubits ($p\LRA s$) using the hybrid state $\rho_{ps}(t)$ in Eq.~(\ref{eq:paevolution}).
Let us first consider teleportation from a polarization qubit
to a single-rail Fock state qubit ($p\RA s$). When
$\ket{B_{1}}_{pp'}$ is
detected in the Bell-state measurement for input state
$\ket{\psi_{\rm t}}_p=a\ket{H}_p+b\ket{V}_p$, the output state can be obtained using Eq.~(\ref{eq:pcmiddle})
as
\begin{align}
\label{eq:psoutcome1}
\rho^{p\RA s}_{B_{1,2}}=&|a|^2\ket{0}_s\bra{0}+|b|^2t^2\ket{1}_s\bra{1}+|b|^2(1-t^2)\ket{0}_s\bra{0}\nonumber\\
&+t(a b^* \ket{0}_s\bra{1}+a^* b \ket{1}_s\bra{0}),
\end{align}
and no unitary transform is required.
If $\ket{B_{2}}_{pp'}$ is measured, a single qubit operation  $(\sigma_z)_s$ is
required to reconstruct $\rho^{p\RA s}_{B_{1,2}}$ in Eq.~(\ref{eq:psoutcome1}).
A phase shifter, described by $\exp[i \varphi a^\dagger a]$ with $\varphi=\pi$, can be used to perform this operation.
The Bell state measurement using linear optics cannot identify $\ket{B_{3}}_{pp'}$ nor $\ket{B_{4}}_{pp'}$.
Furthermore, the $(\sigma_x)_s$ operation required to
implement the bit flip, $|0\rangle\LRA |1\rangle$, is difficult to realize using linear optics.
We thus take only $\ket{B_{1}}_{pp'}$ and $\ket{B_{2}}_{pp'}$ as successful Bell measurement outcomes.
The probability to obtain either of these outcomes is found to be
\begin{align}
&P_{p\RA s}(\theta,\phi)=\nonumber\\
&\mathrm{Tr}\big[(\ket{B_1}_{p p'}\bra{B_1}+\ket{B_2}_{p p'}\bra{B_2})\{\ket{\psi_\mathrm{t}}_p
\bra{\psi_\mathrm{t}}\otimes \rho_{p's}(t;\alpha)\}\big]\nonumber\\
&=t^2/2
\label{eq:spf}
\end{align}
and it is independent of the input state.
The fidelity of state of Eq.~(\ref{eq:psoutcome1}) to the target state $|\psi_t\rangle_s=a\ket{0}_s+b\ket{1}_s$ is
\begin{align}
F_{p\RA s}(a,b)&=_s\langle\psi_t |\rho^{p\RA s}_{B_{1,2}}|\psi_t\rangle_s \nonumber\\
&=|a|^4+|b|^4 t^2+(1-t^2)|a|^2|b|^2+2t|a|^2|b|^2.
\end{align}
The average fidelity is obtained using  Eq.~(\ref{eq:explicitav}) as
\begin{equation}
F_{p\RA s}(t)=\frac{t^2+2t+3}{6}.
\label{eq:Fpa}
\end{equation}

Let us now consider the teleportation in the opposite direction $s\RA
p$. The Bell measurement in the single-rail
Fock state qubit part
 can be performed as follows. After passing through a 50:50 beam splitter,
 two of the Bell states are changed as follows:
$\ket{B_{3}}_{s s'}=2^{-1/2}(\ket{1}_{s}\ket{0}_{s'}+\ket{0}_{s}\ket{1}_{s'})\RA\ket{1}\ket{0}$
and   $\ket{B_{4}}_{s s'}=2^{-1/2}(\ket{1}_{s}\ket{0}_{s'}-\ket{0}_{s}\ket{1}_{s'})\RA\ket{0}\ket{1} $.
As the result, the two Bell states can simply be discriminated using two photodetectors at two output ports of the beam splitter. However, the other two of the Bell states cannot be distinguished using linear optics.
If the outcome of the Bell-state measurement
is $\ket{B_{3}}_{s s'}$ or $\ket{B_{4}}_{s s'}$, the
output state after an appropriate unitary transform is
\begin{align}
\label{eq:rhosp}
\rho_{B_{3,4}}=&\frac{t^4|a|^2+t^2(1-t^2)|b|^2}{4P_3}\ket{H}_p\bra{H}+\frac{ t^2 |b|^2}{4P_3}\ket{V}_p\bra{V}\nonumber\\
&+\frac{t^2(1-t^2)|a|^2+(1-t^2)(2-t^2)|b|^2}{4P_3}\ket{0}_p\bra{0}\nonumber\\
&+\frac{t^3\big(a b^* \ket{H}_p\bra{V}+a^* b \ket{V}_p\bra{H}\big)}{4P_3}.
\end{align}
with success probability $P_{3,4}(\theta,\phi)=((2-t^2)|b|^2+t^2 |a|^2)/4$ obtained in the same way
as Eq.~(\ref{eq:spf}).
The average success probability is found to be $P_{s\RA p}(t)=\langle P_{3,4}(\theta,\phi)\rangle_{\theta,\phi}=1/2$.

The fidelity of $\rho_{B_{3,4}}$ to the target
state $\ket{\psi_t}_p=a\ket{H}_p+b\ket{V}_p$ is straightforwardly obtained as
\begin{align}
\label{eq:Fsp}
F_{s\RA p}(\theta,\phi)=&~_p\bra{\psi_t}\rho_{B_{3,4}}\ket{\psi_t}_p\nonumber\\
=&\frac{t^4|a|^4+t^2(1+2t-t^2)|a|^2|b|^2+t^2 |b|^4}{4P_3}
\end{align}
and the average fidelity over all possible input
states is
\begin{equation}
F_{s\RA p}(t)=t^4 A_1+t^2 A_2+t^2(1+2t-t^2)A_3,
\end{equation}
where
\begin{align}
&A_1=\Big\langle\frac{|a|^4}{4P_3}\Big\rangle_\theta=\frac{c_1^2-4c_1 c_2+3c_2^2+2c_2^2 \log[\frac{c_1}{c_2}]}{2(c_1-c_2)^3},\nonumber\\
&A_2=\Big\langle\frac{|b|^4}{4P_3}\Big\rangle_\theta=\frac{-3c_1^2+4c_1 c_2-c_2^2+2c_1^2 \log[\frac{c_1}{c_2}]}{2(c_1-c_2)^3},\nonumber\\
&A_3=\Big\langle\frac{|a|^2|b|^2}{4P_3}\Big\rangle_\theta=\frac{c_1^2-c_2^2-2c_1c_2 \log[\frac{c_1}{c_2}]}{2(c_1-c_2)^3}
\end{align}
with $c_1=t^2$ and $c_2=2-t^2$.

\begin{figure}[tbp]
\includegraphics[width=180px]{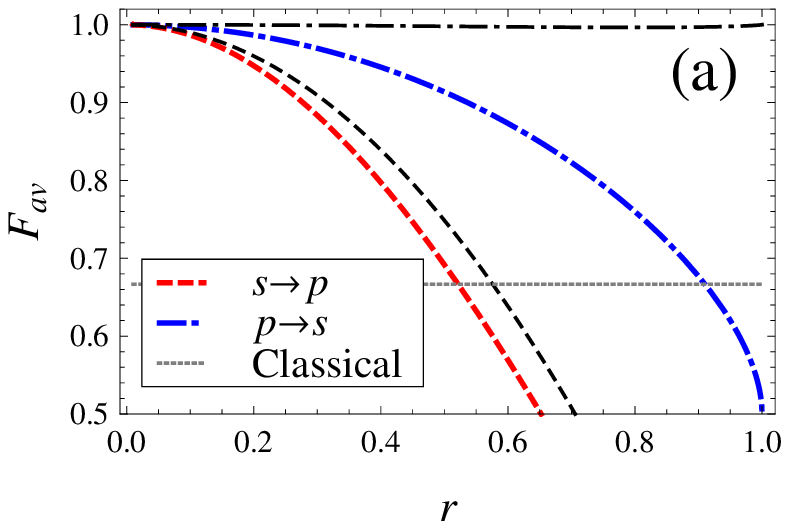}
\includegraphics[width=180px]{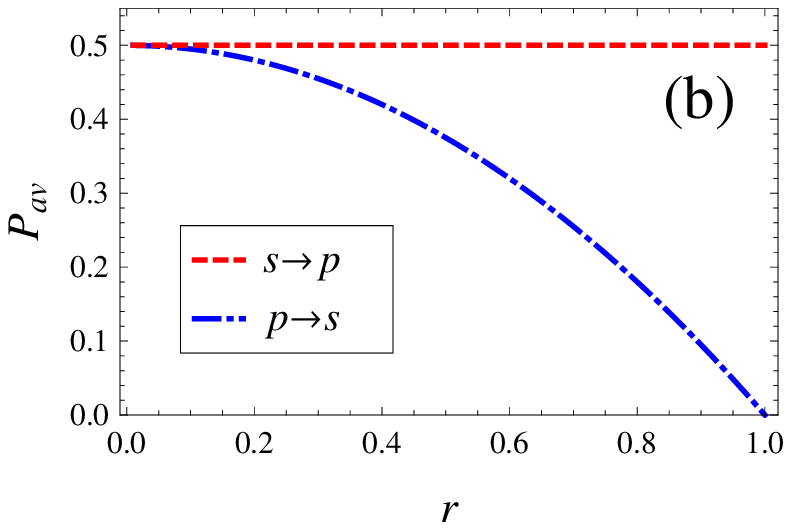}
\caption{ (Color online) (a) Teleportation fidelities of polarization to
single-rail Fock state qubit $p\RA s$ (blue dot-dashed) and the
opposite $s\RA p$ (red dashed). Teleportation fidelities of polarization to coherent-state qubit (black dot-dashed) and the opposite direction (black dashed) are drawn for comparison. (b) Success probability of teleportation of
polarization to single-rail Fock state qubit $p\RA s$ (blue
dot-dashed) and the opposite $s\RA p$ (red dashed).} \label{fig:fpa}
\end{figure}

We plot the teleportation fidelities in Fig.~\ref{fig:fpa}(a) and the
success probabilities in Fig.~\ref{fig:fpa}(b). We observe that the
teleportation fidelity of $p\RA s$ is higher than that of $s\RA p$
because loss in polarization qubit  can be detected
during the Bell-state measurement and discarded, while its success
probability is thus lower as shown in Fig.~\ref{fig:fpa} (b). The
teleportation $s\RA p$  succeeds with probability $1/2$ regardless of $r$
because any decohered single-rail Fock state qubit remains within the original qubit space and
loss is not detected during the Bell-state measurement.
Comparing Figs.~\ref{fig:2DpC} and \ref{fig:fpa}, we observe
that the teleportation fidelity of $p\LRA c$ with small $\alpha$ is higher than that of
$p\LRA s$, although $\rho_{ps}$
contains more entanglement  than $\rho_{pc}$
as shown in Fig.~\ref{fig:negativity}. This can also be understood
as the effect of the basis overlap in coherent-state qubits.

\section{Teleportation with postselection on photon arrival}

We attempt in this section to take into account the effect of postselection on the photon arrival at the receiver's site, while in the previous sections the teleportation fidelity and success probability were calculated considering all of the cases regardless of whether the photon arrived successfully or not (\textit{non-postselected} teleportation).
It was clearly pointed out in Ref.~\cite{KokPRA2000} that considering only the postselected data to calculate the fidelity without resort to operational means of the postselection can be misleading.
In the context of our work, it should be noted that the input qubit to be teleported is an unknown one and should remain unknown after the teleportation for successive use in QIP.
There exists a linear-optical method to implement the quantum-nondemolition detections of single photons that leave the polarization invariant \cite{KokPRA2002}, adopting additional teleportation of the received single polarization qubit through an ideal polarization-entangled state, which reduces the fraction of vacuum state.
We shall now assume that the method in Ref.~\cite{KokPRA2002} is employed to implement postselection for the polarization mode.

\begin{figure}[tbp]
\includegraphics[width=180px]{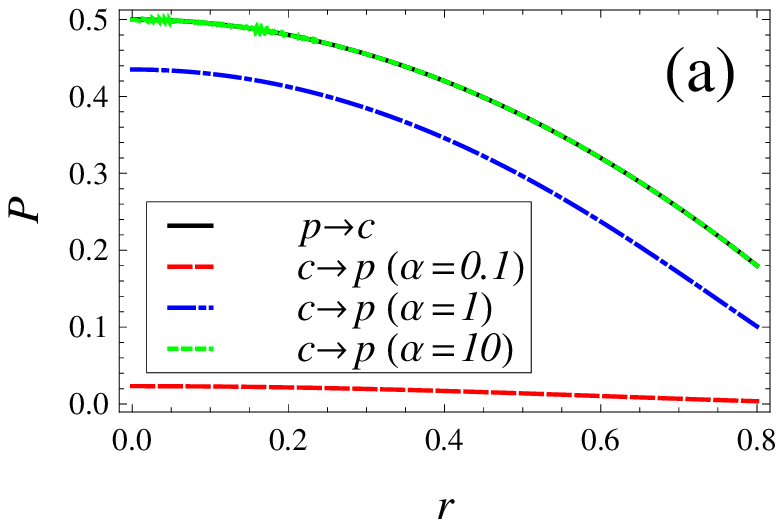}
\includegraphics[width=180px]{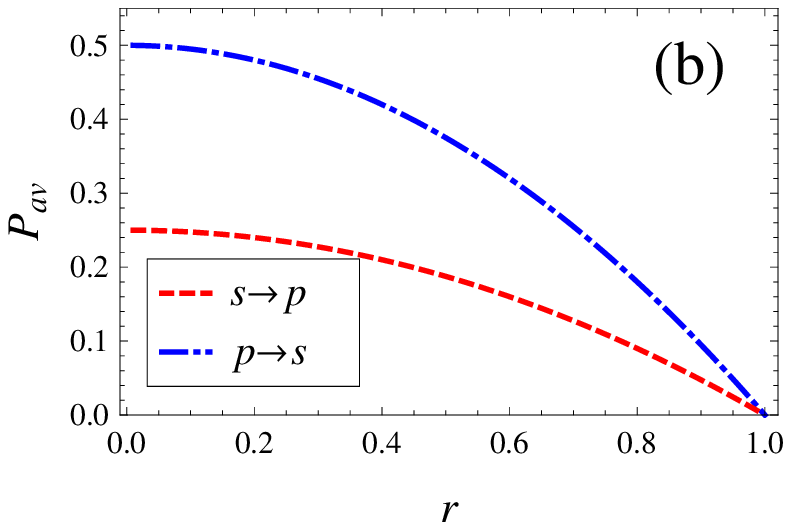}
\caption{(Color online) Success probPpspostabilities after postselection for output qubits in the polarization part for (a) teleportation between polarization and  coherent-state qubits ($p\leftrightarrow c$)
and for (b) teleportation between polarization and  single-rail Fock-state qubits  ($p\leftrightarrow s$).
In (a), the curve for $P_{p\RA c}$ is overlapped with 
 the one for $P_{p\RA c}$ when $\alpha=10$. 
One can see that the success probabilities from the field-like to particle-like qubits are lower than the opposite ones, which is contrary to the cases without postselection for the output qubits in the polarization part.} \label{fig:Ppost}
\end{figure}

We first observe that postselection diminishes the difference between the teleportation fidelities of the two opposite directions by filtering out the vacuum portion in the output state in the polarization mode $p$.
Using the postselection scheme \cite{KokPRA2002}, the output state of teleportation from coherent to polarization qubits  in Eq.~(\ref{eq:Cpout}) is  converted to
\begin{align}
\label{eq:rhopost}
\rho_{c\RA p}^\mathrm{post}=&|a|^2 \ket{H}_p\bra{H}+|b|^2 \ket{V}_p\bra{V}\nonumber\\
&+Q(t) \big(a b^* \ket{H}_p\bra{V}+a^* b \ket{V}_p\bra{H}\big),
\end{align}
thus the coherence terms decrease by the factor $Q(t)=e^{-2\A^2(1-t^2)}$ due to 
the decoherence of the coherent-state part of the channel.
 The average fidelity using Eq.~(\ref{eq:explicitav}) is
$F_{c\RA p}^\mathrm{post}=(2+Q(t))/3$.
When $\A\ll1$, $Q(t)\approx 1$ for all $r$, and $F^\mathrm{post}_{c\RA p}(t)\approx 1$ as well.
The comparison of Eqs.~(\ref{eq:largeA}) and (\ref{eq:rhopost}) shows that the two output forms become identical
for $\A\gg1$ in which  the overlap between coherent states $|\pm\alpha\rangle$ is negligible.

In the case of $s\RA p$, the postselected output state is
\begin{align}
\rho^\mathrm{post}_{s\RA p}=&\frac{t^2|a|^2+(1-t^2)|b|^2}{4P_3}\ket{H}_p\bra{H}+\frac{|b|^2}{4P_3}\ket{V}_p\bra{V}\nonumber\\
&+\frac{t\big(a b^* \ket{H}_p\bra{V}+a^* b \ket{V}_p\bra{H}\big)}{4P_3},
\end{align}
and the average fidelity is changed to
\begin{align}
F^\mathrm{post}_{s\RA p}(t)=t^2 A_1+A_2+(1+2t-t^2)A_3.
\end{align}
When this is compared with $F_{p\RA s}(t)$ in Eq.~(\ref{eq:Fpa}), we find that the two fidelities are again very similar. We find that the largest difference between $F_{p\RA s}(t)$  and $F^\mathrm{post}_{s\RA p}(t)$ is less than $0.01$.

It should also be noted that postselection  decreases the success probability of teleportation into the polarization qubit. The postselection protocol \cite{KokPRA2002} is limited by the polarization Bell measurement
with its success probability 1/2, and failures occur also due to the vacuum part (with a factor of $1-t^2$ for states in Eqs.~(\ref{eq:Cpout}) and (\ref{eq:rhosp})). A factor of $t^2/2$ should thus be multiplied in overall and the postselected probabilities $P^\mathrm{post}_{c,s\RA p}(t)=t^2 P_{c,s\RA p}(t)/2$ are lowered below $P_{p\RA c,s}(t)$, respectively. Probabilities $P^\mathrm{post}_{c\RA p}(t)$ and $P_{p\RA c}(t)$ become identical  only when $\A\gg1$.
We have plotted success probabilities in Fig.~{\ref{fig:Ppost}}.


\section{Remarks}

We have investigated quantum teleportation between two different
types of optical qubits under the effects of decoherence caused by photon losses:
one type is particle-like such as photon-polarization qubit and the other is
field-like such as coherent-state or Fock-state qubits.
The teleportation fidelities and success probabilities depend on
the ``direction'' of teleportation, {\it i.e.}, whether teleportation is performed from one type of qubit to the other or from the latter to the former.

The average fidelity of
teleportation from particle-like to field-like qubits is shown to be
larger than the opposite direction under decoherence. This is due to
the asymmetry of photon losses in the hybrid channel as
well as the possibility of detecting losses in Bell-state measurements.
In the case of teleportation from a single photon qubit using the polarization degree of freedom,
the sender can notice photon loss during the Bell-state measurement in the polarization
qubit part by virtue of its particle-like nature ({\it i.e.}, definite number of particles).
Since the cases with losses are discarded, this enhances the
teleportation fidelity. Even with a teleportation
channel containing very small entanglement, it is possible to obtain
a large teleportation fidelity by filtering the failures in Bell-state
measurements.

The non-orthogonality of the two coherent states that form a qubit basis
is another major factors that
affects the teleportation fidelity.
For example, the
 fidelity of teleportation from polarization to coherent-state qubits with small $\A$ is always higher
than that with large $\A$ due to the larger overlap of the qubit basis
for smaller $\A$.
However,
in order to make a fair comparison, it is important to note that
this non-orthogonality for small $\A$ also increases the classical limit of quantum teleportation.

In terms of success probabilities,
teleportation from field-like to particle-like
qubits shows  higher values.
For example, in the case of teleportation from coherent-state to polarization qubits,
the success probability increases up to 1 as the amplitude of the coherent-state qubit becomes large.

The effect of postselection has been investigated as a trial to increase the fidelity on the particle-like sides. As a result, the fidelities of the teleportation from field-like to particle-like qubits increase and become almost the same to those in the opposite direction. However, the additional resources ({\it i.e.,} preparation of additional polarization entangled states) and the decrease of the success probabilities are the price to be paid.

Our work may provide useful information in the context of
information transfer between systems of different properties.
As an example, since a coherent state
with a large amplitude contains a large number of photons,
the hybrid channel in Eq.~(\ref{eq:pc}) can be considered to be
entanglement between microscopic and macroscopic systems
\cite{W2000,JParkPRA2012,JeongPRL2006,JeongPRA2007,Martini2008,Spagnolo2010,Spagnolo2011}.
Our study may be a framework to study information transfer
between microscopic and macroscopic systems.

\acknowledgments

This work was supported by the National Research
Foundation of Korea (NRF) grant funded by the Korean Government (No.
3348-20100018) and the World Class University program.

\end{document}